\documentclass[%
reprint,
superscriptaddress,
showpacs,
amsmath,amssymb,
prc,
twocolumn,
floatfix, ]%
{revtex4}

\usepackage{color}

\usepackage{graphicx}
\usepackage{dcolumn}
\usepackage{bm}
\usepackage[dvipdfmx,bookmarks=true,colorlinks,%
            citecolor=blue,linkcolor=blue,anchorcolor=blue,filecolor=blue,urlcolor=blue,%
           ]{hyperref}

\allowdisplaybreaks

\newcommand{\beq}{\begin{eqnarray}}
\newcommand{\eeq}{\end{eqnarray}}

\begin{document}


\title{Superdeformed $\Lambda$ hypernuclei from relativistic mean field models}

\author{Bing-Nan Lu}
 \affiliation{State Key Laboratory of Theoretical Physics,
              Institute of Theoretical Physics, Chinese Academy of Sciences,
              Beijing 100190, China}
 \affiliation{Institut f\"ur Kernphysik, Institute for Advanced Simulation, 
              and J\"ulich Center for Hadron Physics, Forschungszentrum J\"ulich, 
              D-52425 J\"ulich, Germany}
\author{Emiko Hiyama}
 \affiliation{Nishina Center, RIKEN, Wako, Saitama, 351-0198, Japan}
\author{Hiroyuki Sagawa}
 \affiliation{Nishina Center, RIKEN, Wako, Saitama, 351-0198, Japan}
\affiliation{Center for Mathematics and Physics, University of Aizu, Aizu-Wakamatsu, 
              Fukushima 965-8580, Japan}
\author{Shan-Gui Zhou}
 \affiliation{State Key Laboratory of Theoretical Physics,
              Institute of Theoretical Physics, Chinese Academy of Sciences,
              Beijing 100190, China}
 \affiliation{Center of Theoretical Nuclear Physics, National Laboratory
              of Heavy Ion Accelerator, Lanzhou 730000, China}

\date{\today}





%


\begin{abstract}

We study the superdeformed (SD) states and corresponding SD hypernuclei of Ar 
isotopes with the multidimensionally-constrained relativistic mean field 
(MDC-RMF) models which can accommodate various shape degree of freedom.
We found that the density profiles of SD states in Ar isotopes show a strong 
localization with a ring structure near the surface, 
while the central part of the density is dilute showing a hole structure.  
This localization of SD density induces an appreciable deformation in 
the hyperon wave function and results in a large overlap between the core and 
the hyperon in the SD hypernuclei of Ar isotopes.
Then the $\Lambda$ separation energy of SD state becomes larger than 
that of normally deformed or spherical ground state.  
This feature is different from that found in other nuclei such as 
$^{32}$S, $^{56}$Ni, and $^{60}$Zn in which the $\Lambda$ separation energy of 
larger deformed state is smaller.   
In this context, the measurement of the $\Lambda$ separation energy may provide 
an important information on the localization of the density profile of SD states.
\end{abstract}
	
\maketitle


\section{Introduction}

A $\Lambda$ particle is free from the nucleon's Pauli exclusion principle in 
nuclei due to the strangeness degree of freedom. 
When a $\Lambda$ particle is added to a nucleus, it lays deeply in 
the nucleus and  
many interesting phenomena were pointed out as the outcome.
For instance, by an addition of a $\Lambda$ particle into the $p$-shell nuclei 
which have an $\alpha$ clustering structure,
there is a dynamical contraction of nuclear size and stabilization in the
binding energy~\cite{Motoba1983_PTP70-189, Motoba1985_PTPSupp81-42, 
Hiyama1996_PRC53-2075, Hiyama1999_PRC59-2351}, 
which is called the glue like role of the $\Lambda$ particle.  
However, the shrinkage effect is dependent on the states in the $p$-shell 
or $sd$-shell nuclei which have both $\alpha$ clustering states and
shell-like compacting states.
Namely, by the addition of a $\Lambda$ particle, we have 20 \% to 30 \% 
nuclear shrinkage in the clustering states, while almost no shrinkage effect 
in shell-like states~\cite{Hiyama1997_PTP97-881}.

As an impurity in normal nuclei, it was pointed out that a hyperon also induces the change of 
the nuclear shape. The nuclear response to the addition of a $\Lambda$ particle 
can be studied by examining the $\Lambda$ separation energy $(S_{\Lambda}$).
The dependence of $S_{\Lambda}$ on the nuclear shape is an interesting topic. 
For example, it has been pointed out that the $\Lambda$ separation energies 
in clustering states of some nuclei, 
e.g., $^{10}_{\Lambda}$Be, $^{13}_{\Lambda}$C, and $^{21}_{\Lambda}$Ne, 
are smaller by 1 to 3 MeV than those in shell-like compact states 
of corresponding hypernuclei from few-body
models~\cite{Hiyama2012_PTP128-105, Zhang2012_NPA881-288} and antisymmetrized 
molecular dynamics (AMD) model~\cite{Isaka2011_PRC83-044323}.

Recently, the self-consistent mean field models, both Skyrme Hartree-Fock 
(SHF) and relativistic mean field (RMF) models, have been applied to study 
the deformation of $p$- and $sd$-shell $\Lambda$ 
hypernuclei~\cite{Zhou2007_PRC76-034312, Win2008_PRC78-054311, 
Schulze2010_PTP123-569, Win2011_PRC83-014301, Lu2011_PRC84-014328}.
In most of the cases, the deformation of core nuclei and the corresponding 
hypernuclei are similar with the same sign.
However, relativistic mean field models predict drastic changes of 
the deformation by the injection of a hyperon, e.g., in $^{13}_{\Lambda}$C 
and $^{29}_{\Lambda}$Si~\cite{Win2008_PRC78-054311, Lu2011_PRC84-014328}.  
Thus in general the polarization effect of hyperons is larger in RMF than 
Skyrme HF (SHF) models~\cite{Schulze2010_PTP123-569}.


Beyond $sd$-shell nuclei, there are many interesting structure issues to study. 
One of them is that in some nuclei there exist superdeformed (SD) states which 
are characterized by the ratio 2:1 between the long and the short deformation axes 
in the coordinate space.   
In nuclei with $A \sim 40$, 
SD states have been observed experimentally~\cite{Svensson2001_PRC63-061301R,
Ideguchi2001_PRL87-222501, OLeary2000_PRC61-064314}
and the structure of these states has been extensively studied by mean field 
models~\cite{Inakura2002_NPA710-261, Bender2003_PRC68-044321, 
Rodriguez-Guzman2004_IJMPE13-139, 
Niksic2006_PRC73-034308, Niksic2006_PRC74-064309, Niksic2009_PRC79-034303}.

A question related to the SD nuclei is, can the SD minima in
normal nuclei persist after the injection of a hyperon? 
If the answer is yes, we have a further question, what happens about the nuclear responses 
in the normally deformed (ND) and SD states by the addition of a hyperon, say, a $\Lambda$?
It is likely that the $\Lambda$ separation energies in SD states are 
smaller than those in ND states because of smaller overlap between the $\Lambda$ 
particle and SD core, as predicted for clustering states by the few-body or AMD 
models~\cite{Hiyama2012_PTP128-105, Zhang2012_NPA881-288, Isaka2011_PRC83-044323}.
Especially, it has been pointed out that the $\Lambda$ separation energies 
in SD states of Ca and Sc isotopes by the addition of a $\Lambda$ particle 
are smaller than those of ND states by Isaka et al.~\cite{Isaka2014_PRC89-024310}.
However, it should be noticed that within the framework of mean field models,
the wave functions of nuclear states might be localized~\cite{Ebran2012_Nature487-341, 
Ebran2013_PRC87-044307}.
It would be interesting to investigate the energy gain by the addition of 
a $\Lambda$ particle to such localized states by using mean field models.

Among many mean field models, in this work we adopt the 
multidimensionally-constrained relativistic mean field (MDC-RMF) models which can accommodate 
various shape degrees of freedom~\cite{Lu2014_PRC89-014323} and have been used 
to study the shape of light hypernuclei~\cite{Lu2011_PRC84-014328} and the
structure of heavy nuclei
~\cite{Lu2012_PRC85-011301R,
Zhao2012_PRC86-057304,
Zhao2014_in-prep}.
This paper is organized as follows.
In Sec.~\ref{sec:theory}, we describe briefly the MDC-RMF model and 
the properties of the $\Lambda$-N interaction employed here.
In Sec.~\ref{sec:results}, the results and discussions are presented. 
A summary is given in Sec.~\ref{sec:summary}.

\section{Theoretical framework~\label{sec:theory}}

Recently multidimensionally-constrained relativistic mean field (MDC-RMF) models were 
developed in order to treat the various shape degrees of freedom in atomic nuclei.
In the MDC-RMF models, the RMF equations are solved in an axially deformed 
harmonic oscillator (ADHO) basis~\cite{Gambhir1990_APNY198-132}. 
The MDC-RMF models have been extended to including the hyperons~\cite{Lu2011_PRC84-014328},
where the Dirac equations for the nucleons and hyperons and the Klein-Gordon 
equations for mesons are solved in the ADHO representation, while the Coulomb 
field is solved by the Green's function method.
Next we mention briefly the formalism of the MDC-RMF models, mostly those 
related to the $\Lambda$ hyperon; 
the readers are referred to Ref.~\cite{Lu2014_PRC89-014323} for more details.

In the MDC-RMF models, the RMF functional can be one of the following four 
forms: the meson exchange or point-coupling nucleon interactions combined 
with the nonlinear or density-dependent couplings.
The deformations $\beta_{\lambda \mu}$ with even $\mu$ can be considered simultaneously.
In the present work we use the meson-exchange type interactions and only 
consider the axial and reflection symmetric shapes.
Thus the projection of the total angular momentum on the symmetric $z$-axis, $\Omega$, 
and the parity $\pi$ are both conserved. 

The starting point of the meson-exchange RMF models for hypernuclei is 
the covariant Lagrangian density
\begin{equation}
 \mathcal{L} = \mathcal{L}_{0} + \mathcal{L}_{\Lambda} ,
 \label{eq:L}
\end{equation}
where $\mathcal{L}_{0}$ is the standard RMF Lagrangian density for the nucleons and
mesons~\cite{Serot1986_ANP16-1, Reinhard1989_RPP52-439, Ring1996_PPNP37-193, 
Bender2003_RMP75-121, Vretenar2005_PR409-101, Meng2006_PPNP57-470, Niksic2011_PPNP66-519} 
and $\mathcal{L}_{\Lambda}$ is that for the hyperon,
\begin{eqnarray}
 \mathcal{L}_{\Lambda} 
 & = & \bar{\psi}_{\Lambda} \left( i \gamma^{\mu} \partial_{\mu} - m_{\Lambda} 
                                 - g_{\sigma\Lambda} \sigma 
                                 - g_{\omega\Lambda} \gamma^{\mu} \omega_{\mu}
                            \right)
             \psi_{\Lambda}
 \nonumber \\
 &   & \mbox{}
 + \frac{f_{\omega\Lambda\Lambda}}{4m_{\Lambda}} 
   \bar{\psi}_{\Lambda} \sigma^{\mu\nu} \Omega_{\mu\nu} \psi_{\Lambda}
 ,
 \label{eq:LL}
\end{eqnarray}
where $m_{\Lambda}$ is the mass of the $\Lambda$ hyperon, 
$g_{\sigma\Lambda}$ and $g_{\omega\Lambda}$ are the coupling constants of 
the $\Lambda$ hyperon with the scalar and vector meson fields, respectively.
The last term represents the tensor coupling between the $\Lambda$ hyperon and 
the $\omega$ field~\cite{Jennings1990_PLB246-325}. 
$\Omega_{\mu\nu}$ is the field tensor of the $\omega$ field defined as 
$\Omega_{\mu\nu}=\partial_{\mu}\omega_{\nu}-\partial_{\nu}\omega_{\mu}$.
Couplings to the $\rho$ meson and the photon vanishes for $\Lambda$ hyperons 
which are neutral and isoscalar. 
In the RMF model, the $\Lambda-\Sigma^0$ mixing, which was important in 
reproducing simultaneously the binding energies of the $s$-shell hypernuclei 
(see, e.g., Refs.~\cite{Akaishi2000_PRL84-3539, Millener2008_NPA804-84}),
is not included explicitly.

Under the mean field approximation, the single particle Dirac equation for 
$\Lambda$ hyperons reads,
\begin{equation}
  \left[ \bm{\alpha} \cdot \bm{p}
       +\beta \left( m_{\Lambda} + S_{\Lambda} \right)
       +V_{\Lambda}
       +T_{\Lambda}
  \right] \psi_{\Lambda i}
  =
  \epsilon_{i} \psi_{\Lambda i}
,
\end{equation}
with the scalar potential $S_{\Lambda}=g_{\sigma\Lambda}\sigma$, 
the vector potential $V_{\Lambda}=g_{\omega\Lambda}\omega$, and the tensor potential,
\begin{equation}
  T_{\Lambda}
  =  - \frac{ f_{\omega\Lambda\Lambda} }{ 2m_{\Lambda} }
    \beta \left( \bm{\alpha} \cdot \bm{p} \right) \omega
.
\end{equation}

The mesons fields $\sigma$, $\omega$, and $\rho$ are obtained by solving 
the Klein-Gordon equations with source terms.
For example, the equation for the $\omega$ meson reads,
\begin{eqnarray}
 \left( -\nabla^{2}+m_{\omega}^{2} \right)\omega 
 & = & 
  \left( g_{\omega}\rho_{B}^{V}+g_{\omega\Lambda}   \rho_{\Lambda}^{V} \right)
 -\frac{ f_{\omega\Lambda\Lambda} }{ 2m_{\Lambda} } \rho_{\Lambda}^{j}
 ,
\end{eqnarray}
where $\rho^V_B$ and $\rho^V_\Lambda$ are the isoscalar-vector densities 
for nucleons and hyperons and 
\begin{eqnarray}
 \rho^{j} & = & 
 i \bm{\partial} \cdot \left( \sum_{i}v_{i}\psi_{i}^{\dagger}\bm{\gamma}\psi_{i} \right)
 ,
\end{eqnarray}
is the divergence of the current.

In this work we use the parameter set PK1~\cite{Long2004_PRC69-034319} 
for the N-N interaction.
The $\Lambda$-N effective interaction is set to be PK1-Y1 which was adjusted to 
reproduce the observed binding energies and spin-orbit splittings of 
the $\Lambda$ hyper nuclei~\cite{Song2010_IJMPE19-2538, Wang2013_CTP60-479}.
In PK1-Y1 the coupling constants for the $\Lambda$-N interaction are 
determined by the relations 
$g_{\sigma \Lambda} = 0.580~g_\sigma$, 
$g_{\omega \Lambda} = 0.620~g_\omega$, and 
$f_{\omega \Lambda \Lambda} = -g_{\omega \Lambda}$.
The spurious motion due to the breaking of the translational invariance is 
treated by including the microscopic center of mass correction
$E_{\rm c.m.} = -\langle P^2 \rangle / (2MA)$ in the total binding 
energy~\cite{Ring1980, Bender2000_EPJA7-467, Long2004_PRC69-034319}.

The pairing effects are included by the BCS approximation with a finite-range 
separable pairing interaction~\cite{Tian2009_PLB676-44, Tian2009_PRC80-024313,
Niksic2010_PRC81-054318},
\begin{equation}
 V( \bm{r}_1 - \bm{r}_2 ) = -G \delta ( \tilde{\bm{R}} - \tilde{\bm{R}}^\prime )
                             P(\tilde{\bm{r}}) P(\tilde{\bm{r}}^\prime)
		 	     \frac{1-\hat{P}_\sigma}{2}  
 ,
\end{equation}
where $G$ is the pairing strength, $\tilde{\bm{R}}$ and $\tilde{\bm{r}}$ are 
the center of mass and relative coordinate between the paired particles, respectively.
$P(\tilde{\bm{r}})$ is a Gaussian shaped function,
\begin{equation}
 P(\tilde{\bm{r}}) = \frac{1}{(4 \pi a^2)^{3/2}} e^{-{\tilde{r}^2}/{a^2}} ,
\end{equation}
where $a$ is the effective range of the pairing force.
In Ref.~\cite{Tian2009_PLB676-44} the strength and range of the separable force 
are adjusted to reproduce the momentum dependency of the pairing gap 
in the nuclear matter calculated from the Gogny forces.
In this work we adopt the parameter set that mimics the Gogny force D1S:
\begin{equation}
 G = 728.0~{\rm{MeV\cdot fm^3}}, \qquad a = 0.644~{\rm{fm}}
 .
\end{equation}

\section{Results and discussions~\label{sec:results}}

\begin{figure}
\includegraphics[width=0.45\textwidth]{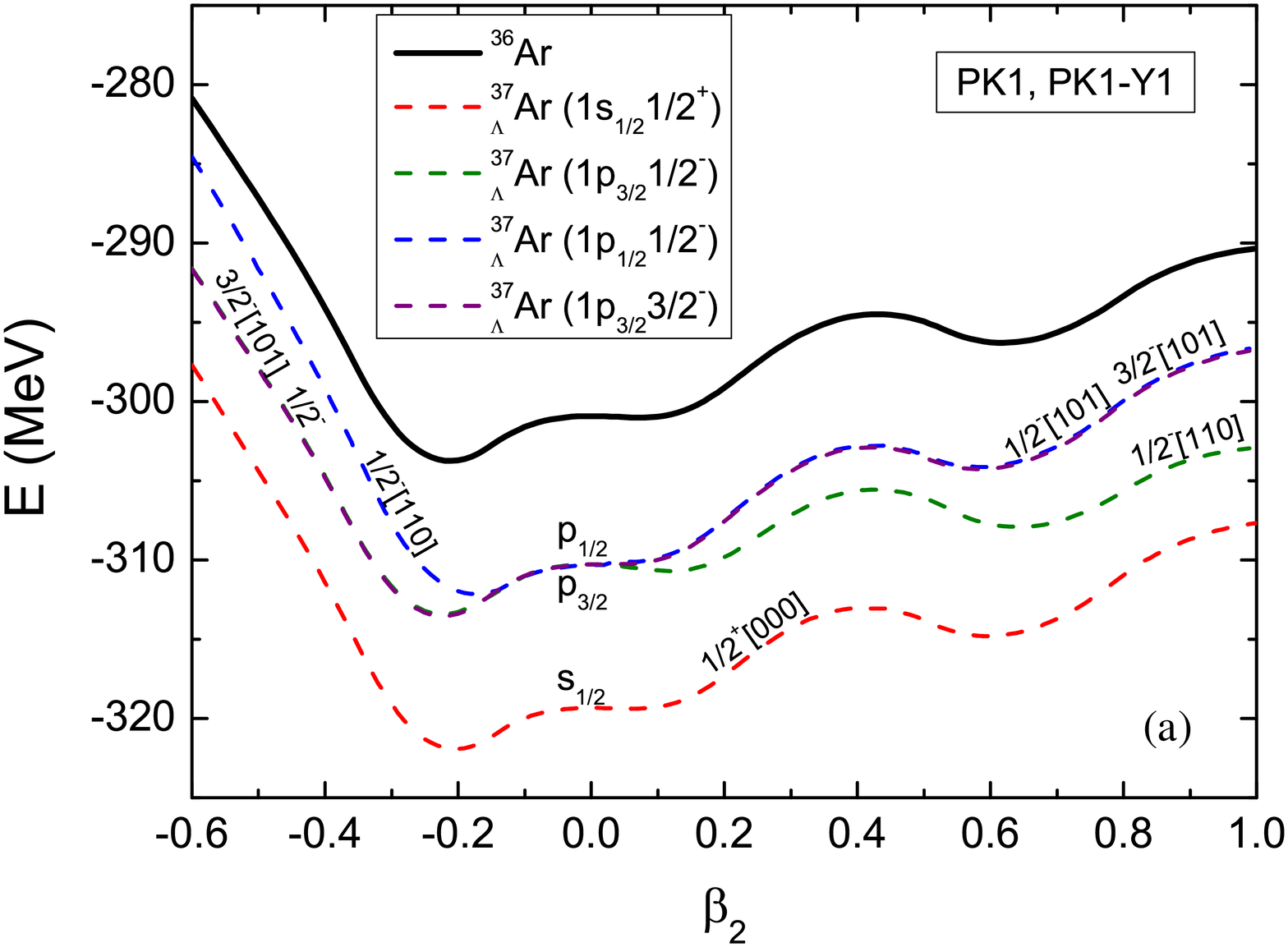}
\includegraphics[width=0.45\textwidth]{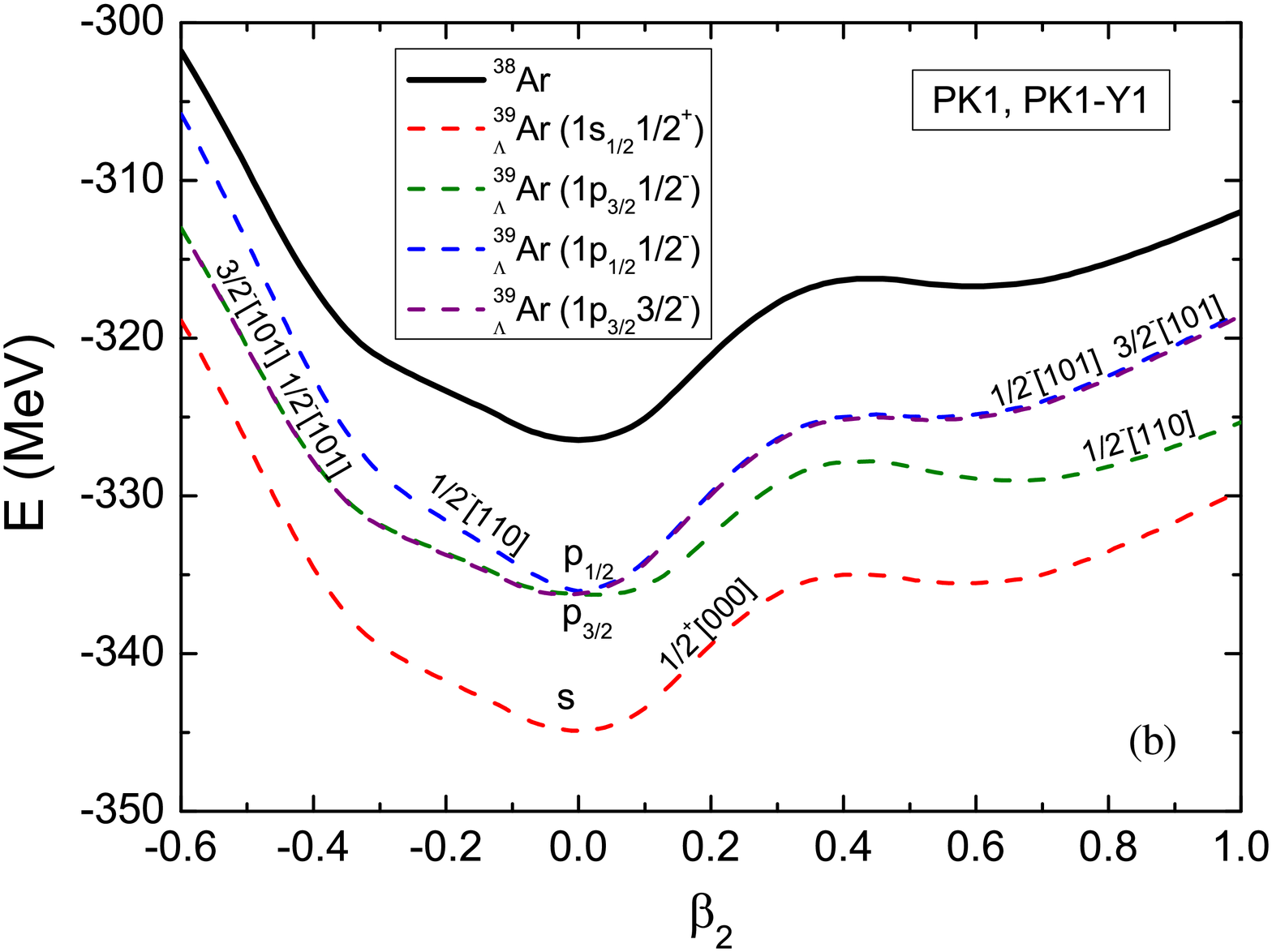}
\includegraphics[width=0.45\textwidth]{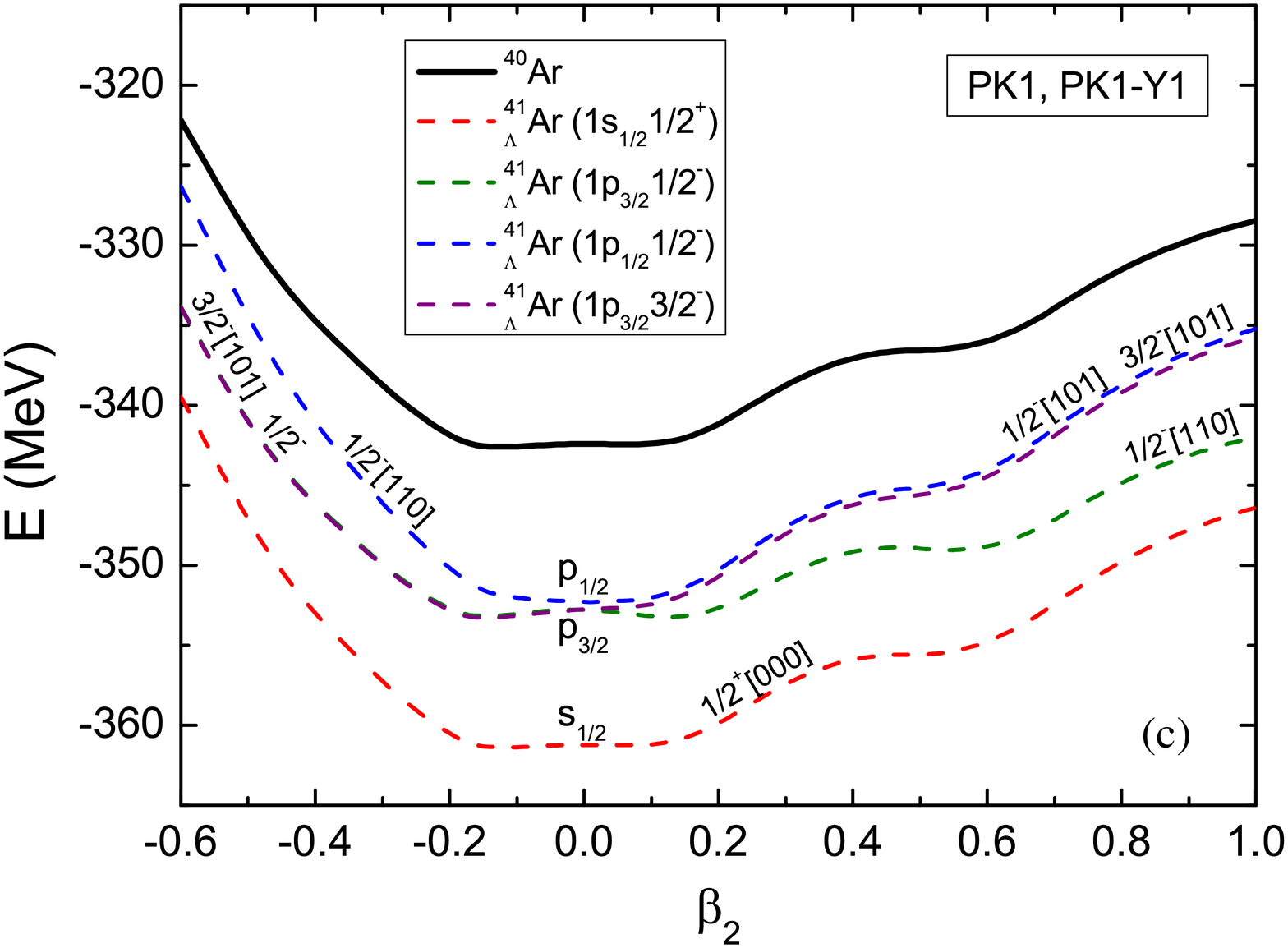}
\caption{~\label{fig:36Ar}
The energy of (a) $^{36}$Ar and $_{\Lambda}^{37}$Ar, 
(b) $^{38}$Ar and $_{\Lambda}^{39}$Ar, and (c) $^{40}$Ar and $_{\Lambda}^{41}$Ar 
as a function of the deformation parameter $\beta_2$, respectively.  
The $\Lambda$ hyperon is injected into the lowest $s$ orbit and three $p$
orbits, respectively. The Nilsson quantum numbers $\Omega^{\pi}[Nn_3m_l]$ of 
the corresponding $\Lambda$ orbits are marked.}
\end{figure}

The calculated energy surfaces of several Ar isotopes ($^{36}$Ar, $^{38}$Ar, and
$^{40}$Ar) and the corresponding $\Lambda$ hypernuclei are shown in Fig.~\ref{fig:36Ar}.
The $\Lambda$ hyperon is injected into the lowest $s$ orbit and three $p$
orbits, respectively. The energy minimum of $^{36}$Ar is found at the oblate 
side with $\beta_2\sim-$0.2. 
The SD minimum is also found at $\beta_2 \sim0.6$.  
The ground state of $^{38}$Ar is spherical as seen 
in Fig.~\ref{fig:36Ar}, while the SD state locates at deformation 
$\beta_2 \sim0.6$ which is similar to that of $^{36}$Ar.   
The energy surface of $^{40}$Ar is rather flat around $\beta_2 \sim 0$
and the ground state is slightly oblate (see Table~\ref{table:Ar-SD}).
Around $\beta_2 \sim0.5$, there is a shallow minimum. 

The energy surfaces of hypernuclei $_{\Lambda}^{A+1}$Ar (a hyperon is put in the 
1$s_{1/2}$ or $\Omega^{\pi}[Nn_3m_l]=1/2^+[000]$ states)   
with $A=36$, 38, and 40 essentially follow those of the core nuclei $^{A}$Ar.
For instance, there are two minima for $_{\Lambda}^{37}$Ar, 
one at the oblate region and the other at the SD region.  
The SD minima become rather shallow in the cases $_{\Lambda}^{39}$Ar and 
$_{\Lambda}^{41}$Ar.  
Because of the small spin-orbit splitting of hyperon-nucleon interaction, 
the 1$p_{1/2}$ and 1$p_{3/2}$ states are almost degenerate in the limit of $\beta_2=0.0$.  
The $1/2^-[110]$ and 3/2$^-$[101] states are splitted in energy by the 
$\beta_2$ deformation in the similar way to the nuclear Nilsson levels 
while the $1/2^-$[101] state is almost degenerate with $3/2^-$[101] state 
because of the very small spin-orbit splitting in the $\Lambda$ potential.  
Around $\beta_2 \sim0.15$, there is a shoulder in the energy surface of $^{36}$Ar.
With one $\Lambda$ added, this shoulder persists; when the $\Lambda$ occupies
the $1/2^-[110]$ state, a very shallow minimum even develops around $\beta_2 \sim0.15$. 
In the following discussions, we will focus on the results in which 
the $\Lambda$ hyperon occupies the $1/2^+[000]$ state.

\begin{table*} 
\caption{~\label{table:Ar-SD}
The quadrupole deformation parameters, root mean square (r.m.s.) radii, 
binding energies, and $\Lambda$ separation energies for the ground states 
and SD states (labelled with asterisks) in several Ar isotopes 
and the corresponding $\Lambda$ hypernuclei. 
The experimental values of binding energies are taken from 
Ref.~\protect\cite{Audi2012_ChinPhysC36-1157, Audi2012_ChinPhysC36-1287,
 Wang2012_ChinPhysC36-1603}.}
\begin{ruledtabular}
\begin{tabular}{lrrrrccccccc}
 {Nucleus} & 
 \multicolumn{4}{c}{Quadrupole deformation parameters} & &
 \multicolumn{2}{c}{R.m.s. radii (fm)} & &
 \multicolumn{3}{c}{Energies (MeV)}
 \tabularnewline
 \cline{2-5} \cline{7-8} \cline{10-12} & 
 $\beta_2$ & $\beta_\mathrm{n}$ & $\beta_\mathrm{p}$ & $\beta_{\Lambda}$  &  & 
 $r_\mathrm{m}$ & $r_{\Lambda}$ &  
 & $E$ & $E_\mathrm{exp}$ & $S_\Lambda$ \tabularnewline
\hline 
$^{36}$Ar           & $-$0.212 & $-$0.208 & $-$0.215 &          &  & 3.238 & 
  &       & $-$303.802 & $-$306.716 &                    \tabularnewline
$_{\Lambda}^{37}$Ar & $-$0.204 & $-$0.205 & $-$0.211 & $-$0.057 &  & 3.220 
  & 2.644 &            & $-$321.979 &          & 18.177   \tabularnewline
$^{38}$Ar           &    0.000 &    0.000 &    0.000 &          &  & 3.281 & 
  &       & $-$326.455 & $-$327.343 &                    \tabularnewline
$_{\Lambda}^{39}$Ar &    0.000 &    0.000 &    0.000 &    0.000 &  & 3.265 &
    2.672 &            & $-$344.896 &          & 18.441   \tabularnewline
$^{40}$Ar           & $-$0.123 & $-$0.112 & $-$0.137 &          &  & 3.338 &
  &       & $-$342.613 & $-$343.810 &                    \tabularnewline
$_{\Lambda}^{41}$Ar & $-$0.117 & $-$0.109 & $-$0.132 & $-$0.037 &  & 3.321 &
    2.698 &            & $-$361.398 &          & 18.785   \tabularnewline
\hline 
$^{36}$Ar*           & 0.620 & 0.610 & 0.630 &       &  & 3.346
 &       & &  $-$296.670 & & \tabularnewline
$_{\Lambda}^{37}$Ar* & 0.597 & 0.599 & 0.619 & 0.172 &  & 3.319
& 2.626 & &  $-$315.194 & & 18.524 \tabularnewline
$^{38}$Ar*           & 0.602 & 0.597 & 0.609 &       &  & 3.403 & 
   & &  $-$317.448 & & \tabularnewline
$_{\Lambda}^{39}$Ar* & 0.589 & 0.596 & 0.603 & 0.187 &  & 3.378 & 
 2.659 &  & $-$336.306 & & 18.858 \tabularnewline
$^{40}$Ar*           & 0.499 & 0.472 & 0.533 &       &  & 3.430 & 
   & &  $-$336.852 & & \tabularnewline
$_{\Lambda}^{41}$Ar* & 0.491 & 0.474 & 0.530 & 0.161 &  & 3.409 
 & 2.685 & &  $-$355.922 & & 19.070 \tabularnewline
\end{tabular}
\end{ruledtabular}
\end{table*}

The calculated deformations of ground states and SD states of
several Ar isotopes are listed in Table~\ref{table:Ar-SD} together with the 
root mean square (r.m.s.) radii, the binding energies, and the $\Lambda$ 
separation energy defined by
\begin{equation}
 S_{\Lambda} = E(^{A+1}_{\Lambda}\mathrm{Ar}) - E(^{A}\mathrm{Ar}).
\end{equation}
From Table~\ref{table:Ar-SD} one finds some familiar features concerning 
the impurity effects of the $\Lambda$ hyperon. 
First, the shrinkage effect appears in all these nuclei studied here as seen
from the values of the r.m.s. radii.
This effect is quite small for the ground states: The radius of a hypernucleus
is smaller by less than 0.02 fm than that of the core nucleus.
Even for the SD states, the change of the radius does not exceed 1\%.
Second, with $A$ increasing, the radius of $\Lambda$ hypernuclei increases,
so does the r.m.s radius of the $\Lambda$ distribution.
Third, the shape of a $\Lambda$ hypernucleus always follows its core nucleus
and the addition of $\Lambda$ does not change the sign of $\beta_2$.
Finally, once the core is deformed, the density distribution of $\Lambda$ is 
also distorted and the deformation parameter $\beta_2$ of a hypernucleus is 
always smaller than that of the corresponding core nucleus.

In Table~\ref{table:Ar-SD}, we find that the $\Lambda$ separation energies of 
Ar isotopes at the ground states have smaller values compared with those of 
the SD states, which is opposite to the case of lighter $\Lambda$ 
hypernuclei such as $^{10}_{\Lambda}$Be~\cite{Zhang2012_NPA881-288, Hiyama2012_PTP128-105}.
The $1/2^+_1$ state in $^9$Be is much more deformed than the ground state $3/2^-_1$. 
When a $\Lambda$ particle is added to these states, the $\Lambda$ separation 
energy in $^9{\rm Be}(1/2^+_1)+\Lambda$ state is smaller than that in   
$^9{\rm Be}(3/2^-_1)+\Lambda$ state. 
This has been explained as that the overlap between nucleons and the $\Lambda$
hyperon in the $1/2^+_1$ state is smaller than that in the ground state.
Next we will show that in Ar isotopes, however, the overlap between the nucleons
and the $\Lambda$ hyperon in the SD state is larger than that in the ground state,
which results in a larger $\Lambda$ separation energy in the SD state.

\begin{table}
\caption{~\label{table:Ca-Zn}
Calculated deformation parameter $\beta_2$, the overlap $I_\mathrm{overlap}$ 
defined in Eq.~(\ref{eq:overlap}), the binding energy $E$, and $\Lambda$ separation 
energy $S_{\Lambda}$ of some $\Lambda$ hypernuclei.
For comparison, the binding energy of the core nucleus $E_{\rm core}$
is also given. The energies are in MeV.
}
\begin{ruledtabular}
\begin{tabular}{lccccc}
 Nucleus& $\beta_{2}$ & $I_\mathrm{overlap}$ & $E_{\rm tot}$ & $E_{\rm core}$ & $S_{\Lambda}$ \\
\hline 
$_{\Lambda}^{37}$Ar  &$-$0.204     & 0.1352 & $-$321.979    & $-$303.802     & 18.177\\
$_{\Lambda}^{37}$Ar* &   0.597     & 0.1370 & $-$315.194    & $-$296.670     & 18.524\\
$_{\Lambda}^{39}$Ar  &   0.000     & 0.1360 & $-$344.896    & $-$326.455     & 18.441\\
$_{\Lambda}^{39}$Ar* &   0.589     & 0.1378 & $-$336.306    & $-$317.448     & 18.858\\
$_{\Lambda}^{41}$Ar  &$-$0.117     & 0.1357 & $-$361.398    & $-$342.613     & 18.785\\
$_{\Lambda}^{41}$Ar* &   0.491     & 0.1378 & $-$355.922    & $-$336.852     & 19.070\\
$_{\Lambda}^{41}$Ca  &   0.00      & 0.1361 & $-$361.422    & $-$342.869     & 18.553\\
$_{\Lambda}^{41}$Ca* &   0.70      & 0.1393 & $-$350.559    & $-$331.317     & 19.242\\
\hline
$_{\Lambda}^{33}$S   &   0.26      & 0.1376 & $-$285.095    & $-$267.002     & 18.093\\
$_{\Lambda}^{33}$S*  &   0.97      & 0.1243 & $-$274.315    & $-$257.951     & 16.364\\
$_{\Lambda}^{57}$Ni  &   0.00      & 0.1461 & $-$506.665    & $-$484.759     & 21.906\\
$_{\Lambda}^{57}$Ni* &   0.40      & 0.1415 & $-$498.610    & $-$477.892     & 20.718\\
$_{\Lambda}^{61}$Zn  &   0.22      & 0.1438 & $-$534.565    & $-$512.924     & 21.641\\
$_{\Lambda}^{61}$Zn* &   0.62      & 0.1415 & $-$527.168    & $-$506.238     & 20.930\\
\end{tabular}
\end{ruledtabular}
\end{table}

To clarify the physical mechanism of the larger $\Lambda$ separation energy in
the SD state, we calculate the overlap $I_\mathrm{overlap}$ of the core 
and the hyperon,
\begin{equation}
 I_\mathrm{overlap} = \int \rho_\mathrm{core}(r,z) \rho_{\Lambda}(r,z) rdr dz ,
 \ \ 
 r = \sqrt{x^2+y^2}
 .
 \label{eq:overlap}
\end{equation}
where the core and the hyperon densities are denoted by $\rho_\mathrm{core}(r,z)$ and
$\rho_{\Lambda}(r,z)$, respectively.

In Table~\ref{table:Ca-Zn}, the overlaps $I_\mathrm{overlap}$ are listed 
together the deformation parameter $\beta_2$, the binding energy $E$, and 
$\Lambda$ separation energy $S_{\Lambda}$ of some $\Lambda$ hypernuclei.
For $_\Lambda^{A+1}$Ar, one finds that the overlap $I_\mathrm{overlap}$ 
in the SD state is always larger than that in the ground state. 
This explains well that the $\Lambda$ separation energy in the SD state
is larger than that of the ground state.
In the following we will examine the density distributions of the core
nucleus and the $\Lambda$ hyperon and show that this new feature actually 
stems from a localized density of nucleons in the core nucleus and 
a stretched density of the $\Lambda$ hyperon in the SD state.  

\begin{figure}
\includegraphics[width=0.48\textwidth]{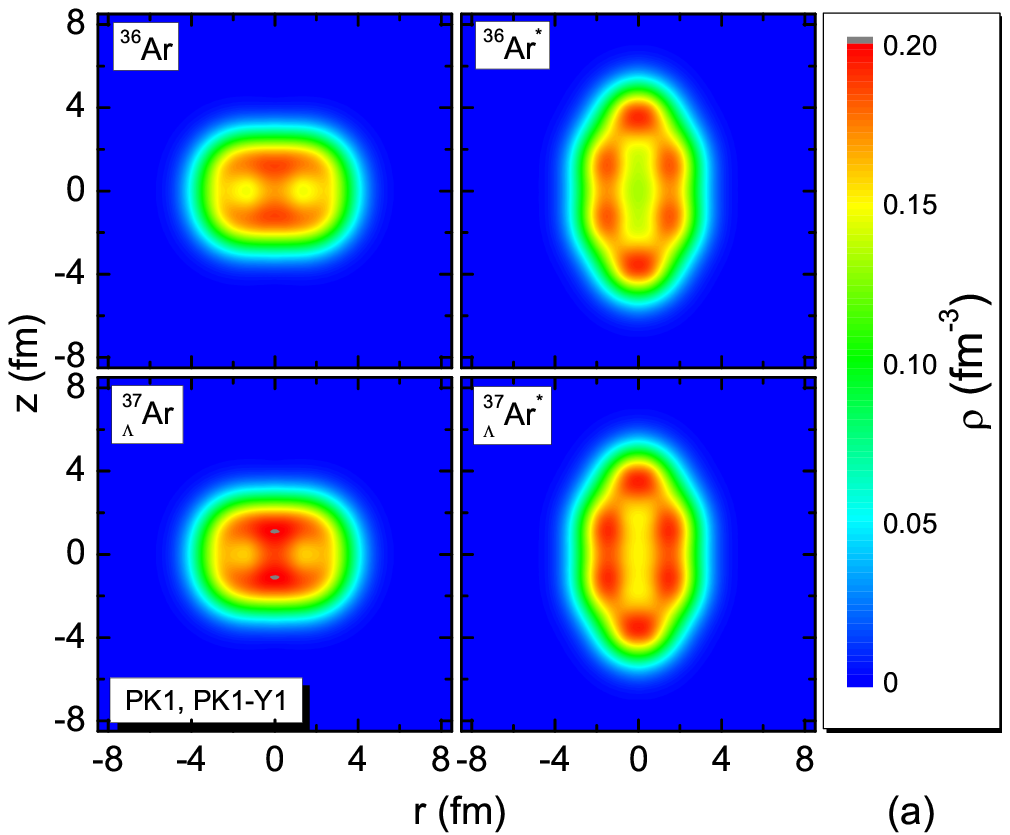}
\includegraphics[width=0.48\textwidth]{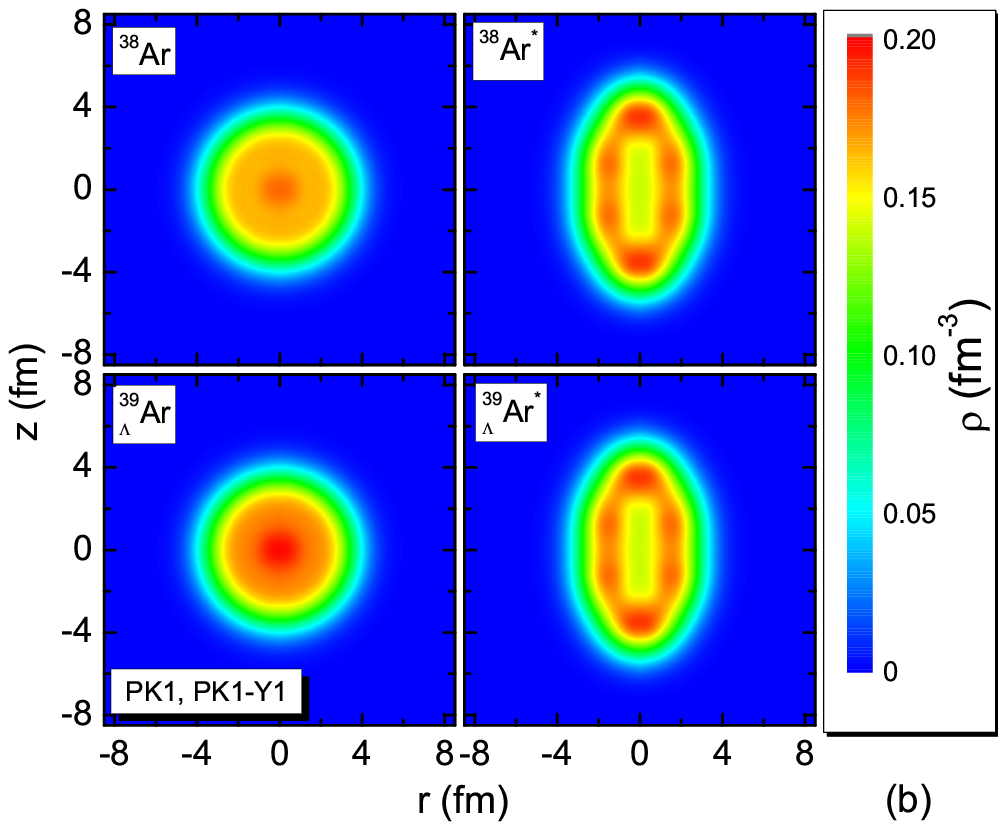}
\includegraphics[width=0.48\textwidth]{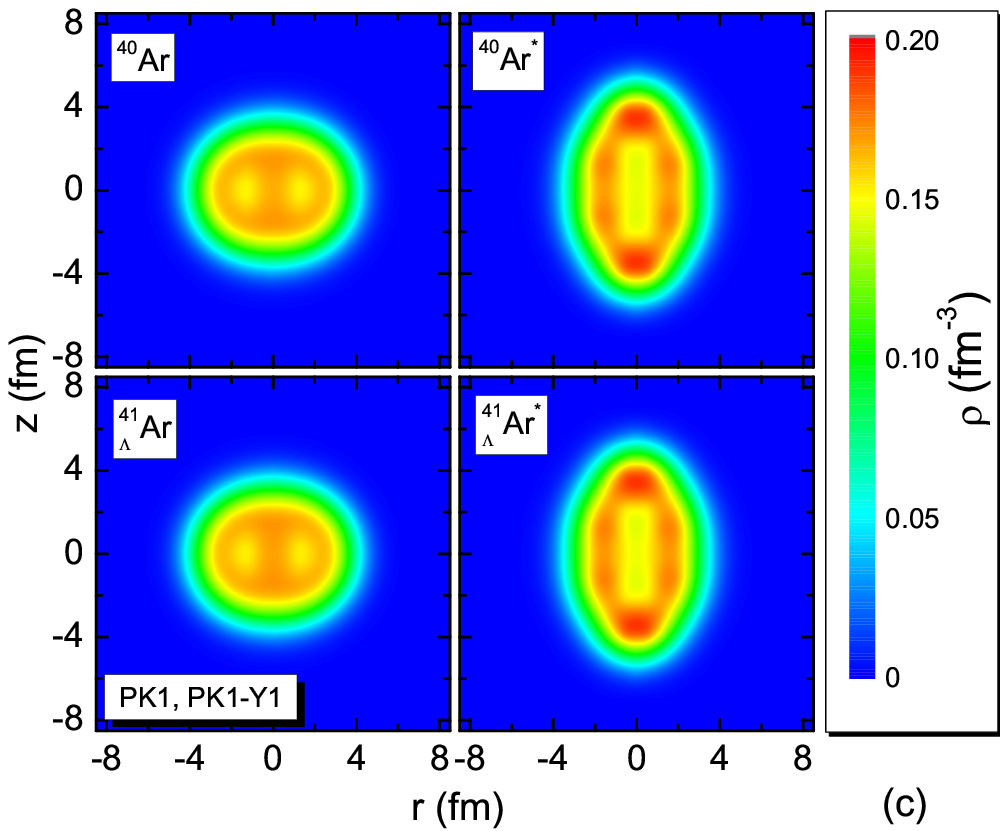}
\caption{~\label{fig:36Ar-2D}
Two-dimensional density distributions in the $r$-$z$ plane 
($r=\sqrt{x^2+y^2}$ and $z$-axis is the symmetric one) for 
(a) $^{36}$Ar and $_{\Lambda}^{37}$Ar and 
(b) $^{38}$Ar and $_{\Lambda}^{39}$Ar 
and (c) $^{40}$Ar and $_{\Lambda}^{41}$Ar 
The asterisks are marked for SD states.  
} 
\end{figure}

We illustrate the calculated two-dimensional density distributions 
in the $r$-$z$ plane (note that the $z$-axis is the symmetric one) 
for the ground and SD states in $^{36,38,40}$Ar and $_{\Lambda}^{37,39,41}$Ar 
in Fig.~\ref{fig:36Ar-2D}.
The ground states of $^{36}$Ar and its hyper-counterpart $_{\Lambda}^{37}$Ar
are clearly oblate.
On the other hand, it is remarkable that the SD density distribution shows 
a clear localization feature with a ring shape near the surface and 
a hole structure with lower density at the center. 
Note that similar localization feature has been found and studied in details 
for the ground state of $^{20}$Ne~\cite{Ebran2012_Nature487-341}.
Furthermore, we see that as increasing number of neutrons, this localization 
with a ring shape of density profile becomes less pronounced as shown in 
Fig.~\ref{fig:36Ar-2D}. This behavior can also be deduced in $\beta_2$ values 
in Table~\ref{table:Ar-SD} where one finds that, as the number of neutrons 
increases, the deformation $\beta_2$ becomes smaller. 
When a $\Lambda$ particle is added to $^{36,38,40}$Ar, the density profiles 
are almost similar with those of corresponding normal nuclei and the localization
effect is still there.
And also, as increasing the number of neutrons, the weakness of localization 
with a ring shape of density profile in hypernuclei is similar with those of 
the corresponding normal nuclei.


\begin{figure}[h]
\includegraphics[width=0.48\textwidth]{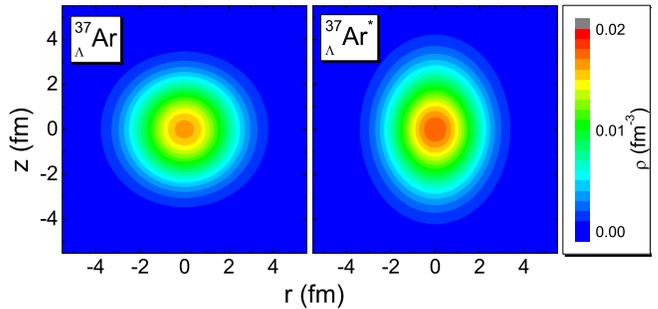}
\caption{~\label{fig:hyp-den}
Two-dimensional density distributions in the $r$-$z$ plane ($r=\sqrt{x^2+y^2}$ 
and $z$-axis is the symmetric one) for the $\Lambda$ hyperon in the ground 
state $_{\Lambda}^{37}$Ar and the SD state $_{\Lambda}^{37}$Ar$^{*}$.
}
\end{figure}

\begin{table}
\caption{~\label{WF-prob}
The probabilities of $\Lambda$ particle wave function in different Nilsson orbits 
for $_{\Lambda}^{37}$Ar in ground state  and SD state, respectively. 
}
\begin{ruledtabular}
\begin{tabular}{ccc}
 Nilsson levels& GS    & SD     \\
\hline 
$1/2^{+}[000]$ & 0.955 & 0.950  \\
$1/2^{+}[220]$ & 0.006 & 0.038  \\
$1/2^{+}[200]$ & 0.034 & 0.006  \\
$1/2^{+}[440]$ & 0.000 & 0.002  \\
\end{tabular}
\end{ruledtabular}
\end{table}

In Fig.~\ref{fig:hyp-den} we show the two-dimensional density distributions 
in the $r$-$z$ plane 
for the $\Lambda$ hyperon in the ground and SD states of $_{\Lambda}^{37}$Ar.
It is seen that in the ground of $_{\Lambda}^{37}$Ar, the $\Lambda$ hyperon
distributes almost isotropically; this is consistent with the small 
$\beta_\Lambda$ value given in Table~\ref{table:Ar-SD}.
Therefore in the ground state of $_\Lambda^{37}$Ar, the overlap between 
the core and the hyperon wave function is small because the core has an 
oblate shape while the hyperon is almost spherical.
However, in the SD state of $_{\Lambda}^{37}$Ar, the density distribution
of the $\Lambda$ hyperon is stretched substantially  along the $z$-axis and the
quadrupole deformation parameter $\beta_\Lambda = 0.172$ (see  Table~\ref{table:Ar-SD}).
The probabilities of the hyperon wave function in different Nilsson orbits 
are given in Table~\ref{WF-prob} for the ground and SD states 
in $_{\Lambda}^{37}$Ar.
It is seen that the hyperon wave function embedded in the SD state has larger 
high-$N$ components (even 4$\hbar\omega$ excitation) than in the ground state. 
These higher-$N$ components with $m_l=0$ corresponds to larger prolate 
shape~\cite{Zhou2010_PRC82-011301R, Li2012_PRC85-024312} and results in 
a larger overlap between the hyperon and the core in $_{\Lambda}^{37}$Ar*.
Since the SD density distribution shows a localization 
feature with a ring shape near the surface and a hole structure with low density 
at the center and the radial coordinate $r$ is included in the integrand,
i.e., $r \rho_\mathrm{core}(r,z) \rho_{\Lambda}(r,z)$ in Eq.~(\ref{eq:overlap}), 
it is natural that the overlap $I_\mathrm{overlap}$ is larger in the SD state
than that in the ground state of $_{\Lambda}^{37}$Ar. 
 
\begin{figure}
\includegraphics[width=0.48\textwidth]{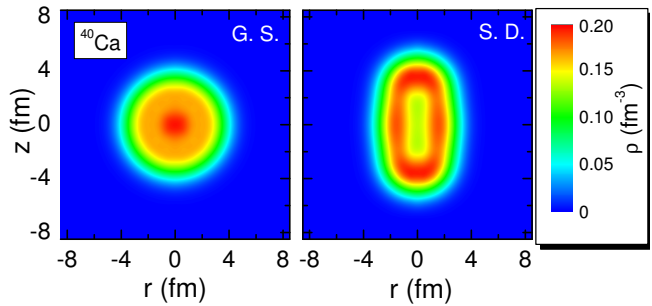}
\caption{~\label{fig:Ca40-2D}
Two-dimensional density distributions in the $r$-$z$ plane 
($r=\sqrt{x^2+y^2}$ and $z$-axis is the symmetric one)  
in the ground state and the SD state of $^{40}$Ca.
}
\end{figure}

To clarify the relation between the large $\Lambda$-separation energy 
in the SD state and the localization feature of the core nucleus, 
we investigate  nuclei which have SD minima in their potential
energy surfaces such as $^{32}$S, $^{40}$Ca, $^{56}$Ni, and $^{60}$Zn and
the corresponding $\Lambda$ hypernuclei.
Among these nuclei, as shown in Fig.~\ref{fig:Ca40-2D}, for $^{40}$Ca 
the SD state has a localization in its density profile with a ring
shape and in other SD states there are no localization effects.
Then it is expected that the localization in the SD state causes 
again a large overlap between the core and the hyperon wave function and,
as a result, the $\Lambda$ separation energy of the SD state becomes larger 
than that of the ground state.
On the other hand, it is likely that since other isotopes, $^{32}$S, 
$^{56}$Ni, and $^{60}$Zn have non-localized core densities,  
the $\Lambda$ separation energies in SD states should be smaller than
those in the ground states.
In Table~\ref{table:Ca-Zn}, we give the overlaps $I_\mathrm{overlap}$ 
together the deformation parameter $\beta_2$, the binding energy $E$, and 
$\Lambda$ separation energy $S_{\Lambda}$ of these $\Lambda$ hypernuclei.
For $_\Lambda^{41}$Ca, the overlap $I_\mathrm{overlap}$ in the SD state is 
indeed larger than that in the ground state, so is $S_{\Lambda}$. 
However, for $_\Lambda^{33}$S, $_\Lambda^{57}$Ni, and $_\Lambda^{61}$Zn,
the overlap between the hyperon and the SD core is smaller than that of 
the ground state which is either normally deformed or spherical.
Then the $\Lambda$ separation energy is larger for the smaller deformed 
state than for the SD state. 

From the above discussions, we conclude that the localization in the SD 
density tends to derive an appreciable deformation in the hyperon wave function 
(see  Fig.~\ref{fig:hyp-den}) and a larger overlap between the core and 
the hyperon which in turn results in a larger $\Lambda$ separation energy.
However, if in the SD state there is no localization in the core density, 
the overlap between the core and the hyperon is smaller in the SD state 
than that in the ground state, which results in a larger $\Lambda$ separation 
energy in the latter.  
These conclusions have been partly demonstrated  in recent AMD model calculations by Isaka et al.:
The density for the SD state in $^{36}$Ar does not show the localization 
feature and the $\Lambda$ separation energy of the ground state is larger 
than that of the SD state in $^{37}_{\Lambda}$Ar~\cite{Isaka2014_in-prep} which 
 is contradictory to the present  results from RMF models.
It would be very interesting to study the localization feature 
of different mean field models for nuclear SD states. 

\section{Summary~\label{sec:summary}} 

We studied the SD states and corresponding SD hypernuclei of Ar isotopes by 
using the RMF model. 
We found that the density profiles of SD states in Ar isotopes show a strong 
localization with a ring structure near the surface, while the central part of 
the density is dilute showing a hole structure.  
The strong localization of the SD state is also found in $^{40}$Ca.   
This localization feature of the SD density induces an appreciable deformation 
in the hyperon wave function of the SD hypernuclei.
Then the $\Lambda$ separation energy of SD state becomes larger than that of 
normally deformed or spherical ground state. 
This feature is different from that found in other nuclei such as 
$_{\Lambda}^{33}$S, $_{\Lambda}^{57}$Ni and $_{\Lambda}^{61}$Zn in which 
the $\Lambda$ separation energy of the SD state is smaller.   
In this context, the measurement of the $\Lambda$ separation energy would  provide 
an important information on the localization of the density profile of SD states.
	
%
%
\begin{acknowledgments}
We thank Dr. M. Kimura and Dr. M. Isaka for helpful discussions.
This work has been partly supported by 
the 973 Program of China (Grant No. 2013CB834400),
the NSF of China (Grants No. 11121403, No. 11120101005, No. 11211120152, and No. 11275248),
the Chinese Academy of Sciences (Grant No. KJCX2-EW-N01),
and
JSPS Grant No. 23224006.
The computation of this work was supported by the High-performance Computing Clusters 
of ITP/CAS and the Supercomputing Center, CNIC of CAS.
\end{acknowledgments}


%

\end{document}